\documentclass[twocolumn,twoside,slac_two]{revtex4}
\usepackage{graphicx}
\usepackage{fancyhdr}
\begin{document}

\title{Sensitivity to SUSY Seesaw Parameters and Lepton
    Flavour Violation}

\author{A.~M.~Teixeira}
\affiliation{Laboratoire de Physique Th\'eorique, UMR 8627,
Universit\'e de
Paris-Sud XI, B\^atiment 201, F-91405 Orsay Cedex, France}
\author{S.~Antusch, E.~Arganda, M.~J.~Herrero}
\affiliation{Departamento de F\'{\i }sica Te\'{o}rica C-XI 
and Instituto de F\'{\i }sica Te\'{o}rica C-XVI, 
Universidad Aut\'{o}noma de Madrid,
Cantoblanco, E-28049 Madrid, Spain}

\begin{abstract}
We address the constraints on the SUSY seesaw parameters arising from 
Lepton Flavour Violation observables. 
Working in the Constrained
Minimal Supersymmetric Standard Model extended 
by three right-handed (s)neutrinos, we 
study the predictions for the branching ratios of $l_j \to
l_i\,\gamma$ and $l_j \to 3\,l_i$ channels. We 
impose compatibility with neutrino  
data, electric dipole moment bounds, and further require
a successful baryon asymmetry
of the Universe (via thermal leptogenesis).
We emphasise the interesting interplay between $\theta_{13}$ and the 
LFV muon decays, pointing out the hints on the SUSY seesaw 
parameters that can arise from measurements of $\theta_{13}$ and 
LFV branching ratios. This is a brief summary of the work of
Ref.~\cite{Antusch:2006vw}. 
\end{abstract}

\maketitle

\thispagestyle{fancy}

\section{Introduction}
Supersymmetric (SUSY) extensions of the Standard Model (SM), 
including three right-handed neutrino superfields, are well motivated
models which can accommodate a seesaw mechanism~\cite{seesaw:I},
and at the same time stabilise the hierarchy between the scale of new
physics and the electroweak (EW) scale. 
One of the most striking phenomenological 
implications of SUSY seesaw models
is the prediction of sizable rates for lepton flavour violating (LFV)
processes~\cite{Borzumati:1986qx},   
many orders of magnitude larger than those expected from the SM
seesaw. In this sense, the
$l_j \to l_i\,\gamma$ and $l_j \to 3\,l_i$ ($i \neq j)$ 
lepton decay channels, as well as $\mu - e$ conversion in
heavy nuclei, are among the most interesting
processes. Experimentally, the
most promising decay is the $\mu \to e\, \gamma$ process, which
exhibits the most stringent present bounds, and offers a
significant improvement regarding the future sensitivity. 

Given the fact that both light and heavy neutrinos enter in the
determination of the LFV rates (via the Yukawa interactions),
a powerful link between the low- and 
high-energy neutrino parameters can be obtained from these LFV processes. 
From the requirement of compatibility with current LFV bounds and with 
low-energy neutrino data, one can then 
extract information on the heavy neutrino sector, thus providing an
indirect access to the heavy neutrino parameters.  

In Ref.~\cite{Antusch:2006vw}, we have systematically explored the 
sensitivity of LFV processes to $\theta_{13}$ in a broad class of SUSY seesaw
scenarios, with different possibilities for the mixing in the neutrino sector. 
We have also incorporated in our analysis the requirement of
generating a successful baryon asymmetry of the Universe (BAU) via
thermal leptogenesis~\cite{Fukugita:1986hr}.
In particular, we have shown that various of the $l_j \to l_i\,\gamma$
and $l_j \to 3\,l_i$ ($i \neq j)$ channels indeed offer 
interesting expectations regarding the sensitivity to  $\theta_{13}$.
This sensitivity to $\theta_{13}$ had been previously pointed
out~\cite{Masiero:2004js,Arganda:2005ji}, for some specific seesaw cases.

Ultimately, and as shown in~\cite{Antusch:2006vw}, the impact of a 
potential $\theta_{13}$ measurement on the LFV branching ratios,
together with the current and future experimental bounds (measurements)
on the latter ratios, may lead to a better knowledge (determination) of 
the heavy neutrino parameters.

\section{LFV within the SUSY Seesaw}
The leptonic superpotential containing the relevant terms to describe a 
type-I SUSY seesaw is given by
$W\,=\,\hat N^c\,Y_\nu\,\hat L \, \hat H_2 \,+\,
\hat E^c\,Y_l\,\hat L \, \hat H_1 \,+\,
\frac{1}{2}\,\hat N^c\,m_N\,\hat N^c\,,$
where $\hat N^c$ is the additional superfield that contains the right-handed
neutrinos and their scalar partners, $Y_{l,\nu}$ are the lepton Yukawa
couplings and $m_N$ is Majorana mass. Henceforth
we will assume that we are in a basis where $Y_l$ and $m_M$ are
diagonal in flavour space.
After EW symmetry breaking, the full $6 \times 6$
neutrino mass matrix is given in terms of the $3 \times 3$ Majorana mass
matrix $m_N$, and the $3 \times 3$ Dirac mass
matrix $m_D\,=\,Y_\nu\,v_2$, where  $Y_{\nu}$ denotes the neutrino Yukawa
couplings and $v_{1(2)}$ 
are the vacuum expectation values of the neutral Higgs
scalars, with $v_{1(2)}= \,v\,\cos (\sin) \beta$ ($v=174$ GeV).

In the seesaw limit, $v \ll m_N$, we obtain the seesaw equation
for the light neutrino masses,
$m_\nu= - m_D^T m_M^{-1} m_D$. The diagonalisation of the full
neutrino mass matrix leads to the six physical Majorana states: three light 
$\nu_i$ and three heavy states $N_i$. Their masses are given by 
$m_{\nu}^\mathrm{diag}=U_\mathrm{MNS}^T \,m_{\nu}\, U_\mathrm{MNS} 
\,=\, \mathrm{diag}\,(m_{\nu_1},m_{\nu_2},m_{\nu_3})$ 
and 
$m_N^\mathrm{diag}\,=\, \mathrm{diag}\,(m_{N_1},m_{N_2},m_{N_3})$.
We use the standard parameterisation for the 
Maki-Nakagawa-Sakata unitary matrix
$U_{\text{MNS}}$~\cite{Umns}, in terms of three mixing angles
$\theta_{12},\theta_{23}$ and $\theta_{13}$, 
and three CP violating phases, $\delta, \phi_1$ and $\phi_2$. 

Following the parameterisation proposed in~\cite{Casas:2001sr}, 
the solution to the seesaw equation can be written as 
\begin{equation}\label{seesaw:casas}
m_D\,=\, i \sqrt{m^{\text{diag}}_N}\, R \,
\sqrt{m^{\text{diag}}_\nu}\,  U^\dagger_{\text{MNS}}\,,
\end{equation}
where $R$ is a generic complex orthogonal $3 \times 3$ matrix, defined by three
complex angles $\theta_i$. This parameterisation allows to accommodate
the experimental data, while leaving room for extra neutrino mixings,
in addition to those in $U_{\text{MNS}}$. It further shows how large
Yukawa couplings $Y_\nu \sim \mathcal{O}(1)$ can be obtained by
choosing large entries in $m^{\text{diag}}_N$.

In our analysis, we have considered scenarios of
hierarchical heavy and light neutrinos, 
$m_{N_1} \ll m_{N_2} \ll m_{N_3}$ and $m_{\nu_1} \ll m_{\nu_2}\ll
m_{\nu_3}$, with 
$m_{\nu_2}^2= \Delta m_{\text{sol}}^2  +  m_{\nu_1}^2$ and 
$m_{\nu_3}^2= \Delta m_{\text{atm}}^2  + m_{\nu_1}^2$. 
Regarding the numerical estimates, we have used
$\Delta m^2_{\text{sol}} =8\times 10^{-5} {\mathrm{eV}}^2$, 
$\Delta m^2_{\text{atm}} =2.5\times 10^{-3} {\mathrm{eV}}^2$,
$\theta_{12}=30^\circ$, $\theta_{23}=45^\circ$, 
$\theta_{13}\lesssim 10^\circ$. For simplicity we have further set
$\delta= \phi_1= \phi_2= 0$.

Within the context of the Constrained
Minimal Supersymmetric Standard Model (CMSSM), universality of the soft SUSY
breaking parameters is imposed at a high-energy scale $M_X$, which we
choose to be the $SU(2)-U(1)$ gauge coupling unification scale ($M_X
\approx 2 \times 10^{16}$ GeV). Instead of 
scanning over the full CMSSM parameter space
(generated by $M_{1/2},\,M_0,\,A_0,\,\tan 
\beta,\,\text{sign} \mu$), we considered 
specific choices for the latter parameters, given by some of the 
``Snowmass Points and Slopes'' (SPS)~\cite{Allanach:2002nj} 
cases defined in Table~\ref{SPS:def:15}.

\begin{table}[h]
\begin{center}
\caption{Values of $M_{1/2}$, $M_0$, $A_0$, $\tan \beta$, 
and sign($\mu$) for the SPS points considered in the analysis.}
\begin{tabular}{|c|c|c|c|c|c|}
\hline 
SPS & $M_{1/2}$ (GeV) & $M_0$ (GeV) & $A_0$ (GeV) & $\tan \beta$ & 
 $\mu$ \\\hline
 1\,a & 250 & 100 & -100 & 10 &  $>\,0 $ \\
 1\,b & 400 & 200 & 0 & 30 &   $>\,0 $ \\
 2 &  300 & 1450 & 0 & 10 &  $>\,0 $ \\
 3 &  400 & 90 & 0 & 10 &    $>\,0 $\\
 4 &  300 & 400 & 0 & 50 &   $>\,0 $ \\
 5 &  300 & 150 & -1000 & 5 &   $>\,0 $\\\hline
\end{tabular}
\label{SPS:def:15}
\end{center}
\end{table}

Regarding our computation of the LFV
observables~\cite{Antusch:2006vw}, 
it is important to stress the following points:
\begin{itemize}
\item
It is a full one-loop computation of the branching ratios (BRs), 
i.e., we include all contributing one-loop diagrams with the 
SUSY particles flowing in the loops. For the case  
of $\l_j \to l_i \gamma$, the analytical formulae can be found
in~\cite{Hisano:1995cp,Arganda:2005ji}. Regarding the 
$\l_j \to 3 \l_i$ decays, the complete set of diagrams (including
photon-penguin, $Z$-penguin, Higgs-penguin and box diagrams) 
and formulae are given in~\cite{Arganda:2005ji}.    

\item 
The computation is performed in the physical basis for all SUSY
particles entering in the loops. In other words, we do not
use the Mass Insertion Approximation (MIA).

\item
To obtain the low-energy parameters of the model 
the full renormalisation group equations
(RGEs), including relevant terms and equations for the 
neutrinos and sneutrinos, are firstly run down from $M_X$ to $m_{N}$.
At the seesaw scale (in particular at $m_{N_3}$),
we impose the boundary condition of
Eq.~(\ref{seesaw:casas}). After the decoupling of the heavy neutrinos
and sneutrinos, the new RGEs are then run down from $m_{N_1}$ to 
the EW scale, at which the observables are computed. 
More concretely, we do not use the Leading Log Approximation (LLog),
but rather numerically solve the full one-loop RGEs.

\item
The numerical implementation of the above procedure is achieved by
means of the public Fortran code {\tt
SPheno2.2.2}~\cite{Porod:2003um}, which has been adapted 
in order to fully incorporate the right-handed
neutrino (and sneutrino) sectors, as well as the full lepton flavour 
structure~\cite{Arganda:2005ji}. 

\item 
The SPheno code has been further enlarged by additional
subroutines that compute the LFV branching ratios for all the $\l_j
\to l_i \gamma$ and $\l_j \to 3 \l_i$ channels~\cite{Arganda:2005ji}. 
We have also included subroutines~\cite{Antusch:2006vw} to implement 
the requirement of successful baryogenesis 
(which we define as having $n_B/n_\gamma \in [10^{-10},10^{-9}]$) 
via thermal leptogenesis in the presence of upper bounds on 
the reheat temperature, and to ensure
compatibility with present bounds on lepton
electric dipole moments (EDMs): {$\mbox{EDM}_{e \mu \tau}$} {$\lesssim (6.9
\times 10^{-28}, 3.7 \times 10^{-19}, 4.5 \times 10^{-17}) \,
\mbox{e.cm}$}~\cite{Yao:2006px}. 

\end{itemize}

In what follows we present our main results for the case of
hierarchical heavy neutrinos. We also include a comparison with
present bounds on LFV
rates~\cite{Brooks:1999pu,Aubert:2005wa,Aubert:2005ye,Bellgardt:1987du,Aubert:2003pc} and their future
sensitivities~\cite{mue:Ritt,Akeroyd:2004mj,Iijima,Aysto:2001zs}
collected in Table~\ref{LFV:bounds:future}.

\begin{table}[h]
\begin{center}
\caption{Present bounds and future sensitivities for the LFV 
processes.}
\begin{tabular}{|c|c  c |}
\hline
LFV process & Present bound & Future sensitivity \\
\hline
BR($\mu \to e\,\gamma$) & $1.2 \times 10^{-11}$  & $1.3 \times
10^{-13}$  \\
BR($\tau \to e \,\gamma$) & $1.1 \times 10^{-7}$ & 
 $10^{-8}$ \\
BR($\tau \to \mu \,\gamma$) & $6.8 \times 10^{-8}$  &
$10^{-8}$  \\
BR($\mu \to 3\,e$) & $1.0 \times 10^{-12}$  & 
$10^{-13}$  \\
BR($\tau \to 3\,e$) & $2.0 \times 10^{-7}$  & 
$10^{-8}$  \\
BR($\tau \to 3\,\mu$) & $1.9 \times 10^{-7}$  & 
$10^{-8}$  \\\hline
\end{tabular}
\label{LFV:bounds:future}
\end{center}
\end{table}

\section{Results and Discussion}
\begin{figure*}[htb]
\centering\hspace*{-10mm}
\begin{tabular}{cc}
\includegraphics[width=56mm,angle=270]{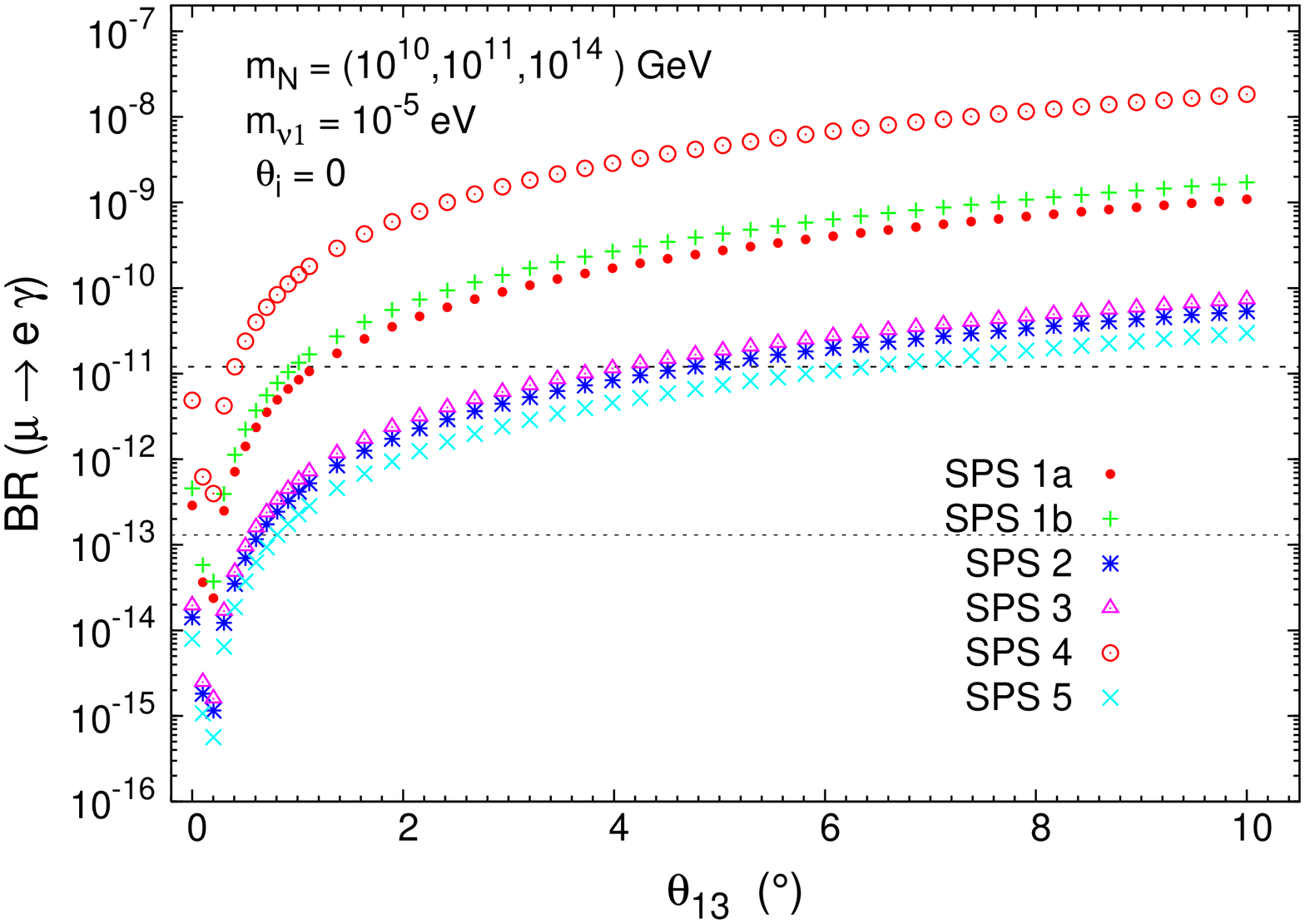}\hspace*{3mm}
&
\includegraphics[width=56mm,angle=270]{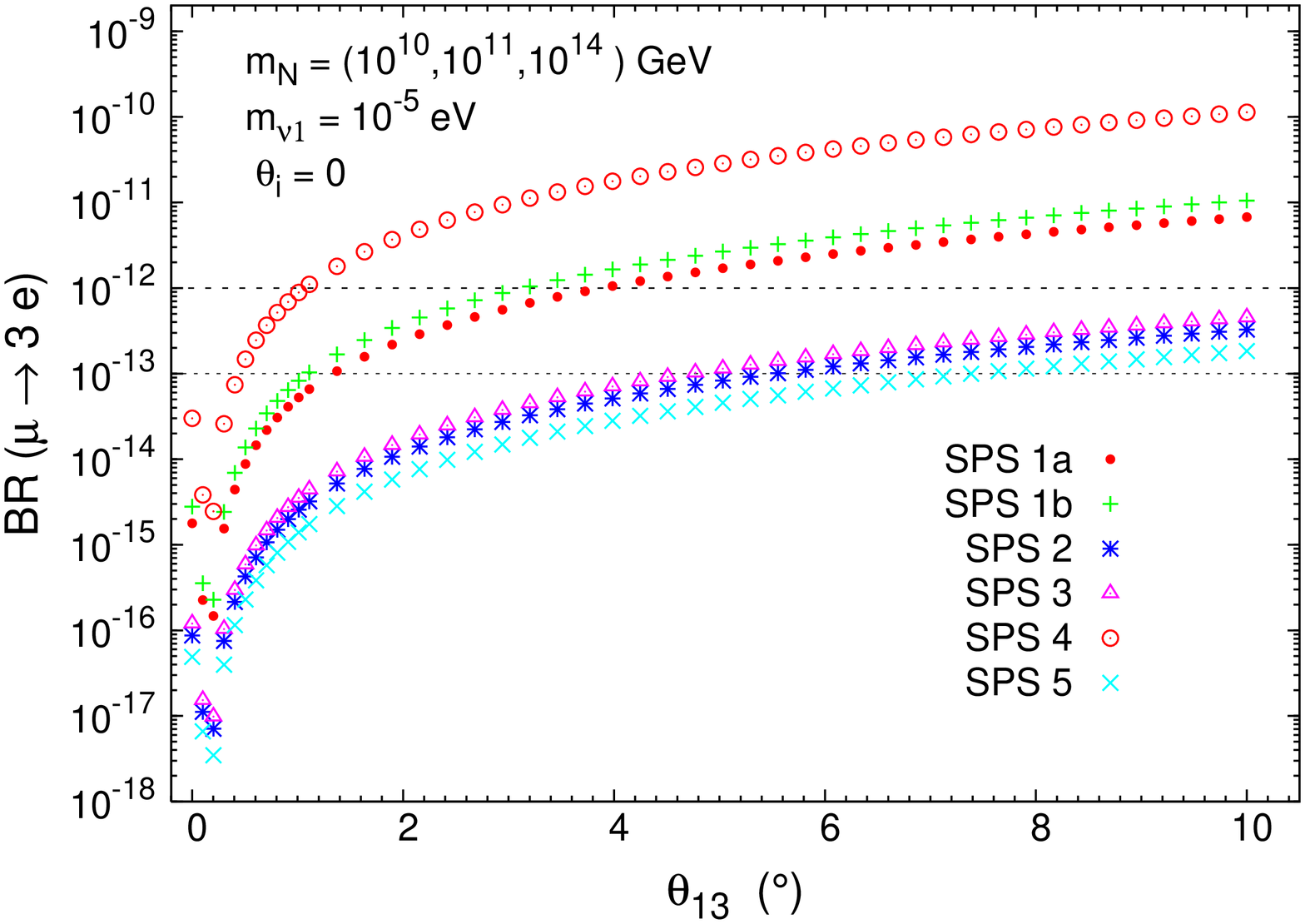}
\end{tabular}
\caption{BR($\mu \to e\, \gamma$) and BR($\mu \to 3\,e $)
  as a function of $\theta_{13}$ (in degrees), for SPS 1a
  (dots), 1b (crosses), 2 (asterisks), 3 (triangles), 4 (circles) and 5
  (times). A dashed (dotted)
  horizontal line denotes the present experimental bound (future
  sensitivity).} \label{fig:SPS:t13:ad}
\end{figure*}

Here we focus on the sensitivity
of the BRs to $\theta_{13}$, and on the dependence on other relevant
parameters, which, for the case of
hierarchical heavy neutrinos, are the heaviest 
mass $m_{N_3}$, $\tan\beta$, $\theta_1$ and $\theta_2$ (using the $R$
parameterisation of~\cite{Casas:2001sr}). The other input seesaw 
parameters $m_{N_1}$, $m_{N_2}$ and $\theta_3$, play a secondary role
since the BRs do not strongly depend on them. 
Finally, we comment on the hints on the SUSY seesaw parameters
that can be derived from a measurement of the BRs and $\theta_{13}$.

For $R = 1$, the predictions of the BRs as functions of 
$\theta_{13}$ in the experimentally allowed range of 
$\theta_{13}$, $0^\circ \leq
\theta_{13} \leq 10^\circ$ are illustrated in Fig.~\ref{fig:SPS:t13:ad}. 
In this figure we also include the present and
future experimental sensitivities for the channels. We clearly see 
that the BRs of  $\mu \to e \gamma$ and $\mu \to 3 e$
are extremely sensitive to $\theta_{13}$, with 
their predicted rates varying many orders of magnitude along the
explored $\theta_{13}$ interval. The BRs of 
$\tau \to e\, \gamma$ and $\tau \to 3\,e$ channels are also 
sensitive to $\theta_{13}$, but
experimentally less challenging. The other LFV channels,
$\tau \to \mu \gamma$ and  
$\tau \to 3 \mu$, are nearly insensitive to this
parameter (see~\cite{Antusch:2006vw}). 
In the  case of $\mu \to e \gamma$ this strong sensitivity was previously 
pointed out in Ref.~\cite{Masiero:2004js}. In~\cite{Arganda:2005ji},
working within a full RGE approach, it was noticed that 
$\mu \to e \gamma$ and $\mu \to 3 e $
were the channels that, in addition to manifesting a clear
$\theta_{13}$ dependency, were the most promising from the
experimental detection point of view. 

The most important conclusion from Fig.~\ref{fig:SPS:t13:ad} is
that, for this choice of parameters, the predicted BRs for both 
muon decay channels, $\mu \to e \gamma$ and
$\mu \to 3 e$, are clearly within the present experimental reach for
several  of the studied SPS points. The most stringent channel is
manifestly $\mu \to e \gamma$ where the 
predicted BRs for all the SPS points are clearly above the present
experimental bound  
for $\theta_{13} \gtrsim 5^\circ$. With the expected improvement in
the experimental 
sensitivity to this channel, this would happen for $\theta_{13}
\gtrsim 1^\circ$. 

\medskip
In addition to the small neutrino mass generation, the seesaw
mechanism offers the interesting possibility of 
baryogenesis via leptogenesis~\cite{Fukugita:1986hr}. Thermal leptogenesis 
is an attractive and minimal mechanism to produce a successful BAU, even
compatible with present data, $n_\mathrm{B} /n_\gamma 
\,\approx\, (6.10\,\pm\,0.21)\,\times\,10^{-10}$~\cite{Spergel:2006hy}. 
In the supersymmetric version of the seesaw mechanism, it can
be successfully implemented provided that the 
following conditions can be satisfied. Firstly, Big Bang
Nucleosynthesis gravitino  
problems have to be avoided, which is possible, for instance, for 
sufficiently heavy gravitinos. Since we consider the gravitino mass 
as a free parameter, this condition can be easily achieved. In any
case, further  
bounds on the reheat temperature, $T_\mathrm{RH}$, still arise from 
decays of gravitinos into the lightest supersymmetric particle (LSP). In the
case of heavy gravitinos and neutralino LSP masses in the range 100-150
GeV (which is the case of our work), one obtains 
$T_\mathrm{RH} \lesssim 2 \times 10^{10}$ GeV.
In the presence of these constraints on 
$T_\mathrm{RH}$, the favoured region by thermal leptogenesis 
corresponds to small (but non-vanishing) complex $R$-matrix angles 
$\theta_i$. For vanishing $U_\mathrm{MNS}$ CP phases the constraints on $R$ are
basically $|\theta_2|,|\theta_3|  
\lesssim 1 \, \mbox{rad}$ (mod $\pi$). 
Thermal leptogenesis also
constrains $m_{N_1}$ to be roughly in the range $[10^9\:
\mbox{GeV},10 \times T_\mathrm{RH}]$ (see also
\cite{Giudice:2003jh,Antusch:2006gy}). 

In~\cite{Antusch:2006vw}
we have explicitly calculated the produced BAU in the presence of upper bounds
on the reheat temperature $T_\mathrm{RH}$. We have furthermore set as
``favoured BAU values'' those that are within the interval
$[10^{-10},10^{-9}]$, which contains 
the WMAP value, and chosen the value of $m_{N_1} =
10^{10}$ GeV in most of our analysis.
Similar studies of the constraints from leptogenesis on LFV rates have been
done in~\cite{Petcov:2005jh}.

\medskip
For very small values of $m_{\nu_1}$ ($m_{\nu_1}\sim
\mathcal{O}(10^{-5} \,\mathrm{eV})$) a  baryon asymmetry 
in the range $10^{-10}$ to $10^{-9}$ can be obtained for a
considerable region of the $|\theta_2|$ parameter space, with the BRs
exhibiting a clear
sensitivity to the value of $\theta_{13}$~\cite{Antusch:2006vw}. On
the other hand, the situation changes dramatically for larger values
of $m_{\nu_1}$. 

\begin{figure}[t]
\centering\hspace*{-3mm}
\includegraphics[width=60mm,angle=270]{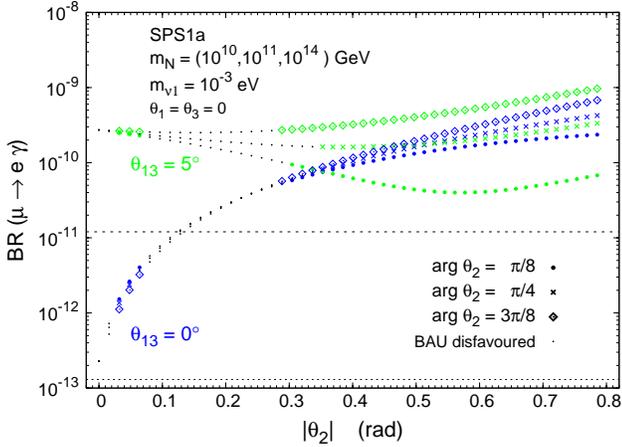}
\caption{BR($\mu \to e\, \gamma$) as a function
      of $|\theta_2|$, for $\arg
      \theta_2\,=\,\{\pi/8\,,\,\pi/4\,,\,3\pi/8\}$ (dots, times,
      diamonds, respectively) and
      $\theta_{13}=0^\circ$, 5$^\circ$ (blue/darker, green/lighter lines).
      We take $m_{\nu_1}\,=\,10^{-3}$ eV.
      In all cases black dots
      represent points associated with a disfavoured BAU scenario and 
      a dashed (dotted) horizontal line denotes the present 
      experimental bound (future sensitivity).} \label{fig:modt2:argt2:1214}
\end{figure}

In Fig.~\ref{fig:modt2:argt2:1214}, we display the dependence of 
the most sensitive BR to $\theta_{13}$, BR$(\mu \to e\, \gamma)$, on
$|\theta_2|$. We consider two
particular values of $\theta_{13}$,
$\theta_{13}=0^\circ\,,5^\circ$ and choose SPS 1a. Motivated from the
thermal leptogenesis favoured
$\theta_2$-regions~\cite{Antusch:2006vw}, we take $0 \,\lesssim \,
|\theta_2|  \,\lesssim \, \pi/4$, with $\arg \theta_2
\,=\,\{\pi/8\,,\,\pi/4\,,\,3\pi/8\}$. We choose  $m_{\nu_1}\,=\,10^{-3}$~eV, 
while for the heavy neutrino masses we 
take $m_{N}\, =\, (10^{10},\,10^{11},\,10^{14})$~GeV.

While for smaller values of $|\theta_2|$
the branching ratio displays a clear sensitivity to having
$\theta_{13}$ equal or different from zero (a separation larger than
two orders of 
magnitude for $|\theta_2| \lesssim 0.05$), the effect of $\theta_{13}$ is
diluted for increasing values of $|\theta_2|$. For $|\theta_2| \gtrsim 0.3$ 
the BR($\mu \to e\, \gamma$)
associated with $\theta_{13}\,=\,5^\circ$ can be even smaller than for 
$\theta_{13}\,=\,0^\circ$. This implies that in this case, 
a potential measurement of BR($\mu \to e\, \gamma$) would not be
sensitive to $\theta_{13}$. Similar results were obtained for
$\theta_3$, but for shortness are not shown here.

Concerning the EDMs, which are clearly non-vanishing in the presence
of complex $\theta_i$, we have checked that all the predicted values for 
the electron, muon
and tau EDMs are well below the experimental bounds.  

\begin{figure}[t]
\centering\hspace*{-3mm}
\includegraphics[width=60mm,angle=270]{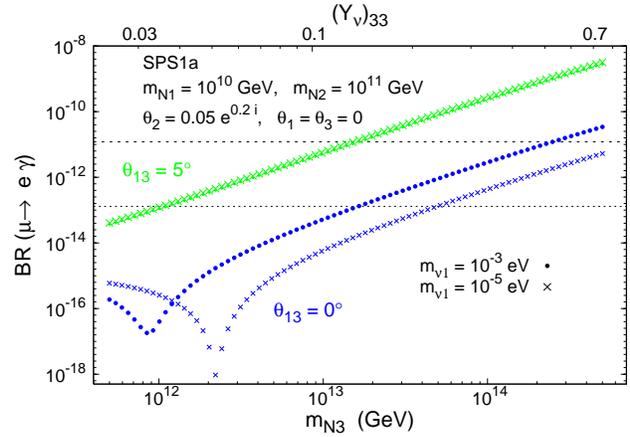}
\caption{BR($\mu \to e\, \gamma$) as a function of 
      $m_{N_3}$ for SPS 1a, with
      $m_{\nu_1}\,=\,10^{-5}$~eV and $m_{\nu_1}\,=\,10^{-3}$~eV (times, dots,
      respectively), and
      $\theta_{13}=0^\circ,\,5^\circ$ (blue/darker, green/lighter lines). 
      Baryogenesis is enabled by the choice 
      $\theta_2\,=0.05\,e^{0.2\,i}$ ($\theta_1=\theta_3=0$). 
      On the upper horizontal axis we display the associated value of
      $(Y_\nu)_{33}$. A dashed (dotted)
      horizontal line denotes the present experimental bound (future
      sensitivity).}  
    \label{fig:MN3:MN2:SPS1a}
\end{figure}

\medskip
We now consider the dependence of BR$(\mu \to e \gamma)$ on
$m_{N_3}$. As displayed in Fig.~\ref{fig:MN3:MN2:SPS1a}, there is a 
strong sensitivity of the BRs to $m_{N_3}$. In fact, the BRs vary by
as much as six orders of magnitude in the explored range 
of $5 \times 10^{11} \, {\rm GeV} \leq  m_{N_3} \leq 5 \times 10^{14}
\, {\rm GeV}$. 
Notice also that for the largest values of $m_{N_3}$ considered, 
the predicted rates for $\mu \to e \gamma$ enter into the present
experimental reach. Although not shown here, it is also worth mentioning 
that by comparing our full results with the LLog predictions, we found
that the LLog approximation dramatically fails in some
cases~\cite{Antusch:2006vw}. Similar effects were also noticed
in~\cite{Petcov:2003zb,Chankowski:2004jc}.

\medskip
Regarding the $\tan \beta$ dependence of the BRs we obtained that 
the BR grow as $\tan^2 \beta$.
In fact, the
hierarchy of the BR predictions for the several SPS points (as already 
manifest in Fig.\ref{fig:SPS:t13:ad}) is dictated by the
corresponding $\tan \beta$ value, with a secondary role being played by the
given SUSY spectra. We found the following generic hierarchy:
BR$_{\rm SPS4}$~$>$~BR$_{\rm SPS1b}$~$\gtrsim$~BR$_{\rm SPS1a}$~
$>$~BR$_{\rm SPS3}$~$\gtrsim$~BR$_{\rm SPS2}$ ~$>$~BR$_{\rm SPS5}$.
\begin{figure*}[t]
\centering\hspace*{-1mm}
\includegraphics[width=85mm]{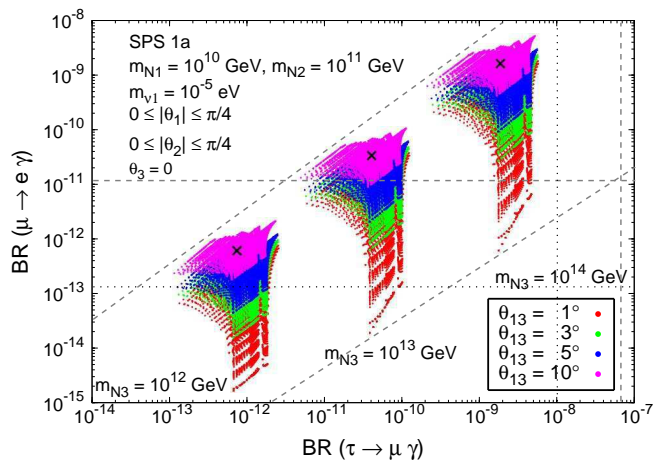}
\caption{Correlation between BR($\mu \to e\,\gamma$) and 
      BR($\tau \to \mu\,\gamma$) as a function of $m_{N_3}$, for SPS
      1a. The areas displayed represent the scan over $\theta_i$ 
      as given in Eq.~(\ref{doubleBR:input}). From bottom to top, 
      the coloured regions correspond to 
      $\theta_{13}=1^\circ$, $3^\circ$, $5^\circ$ and $10^\circ$ (red,
      green, blue and pink, respectively). Horizontal and vertical 
      dashed (dotted) lines denote the experimental bounds (future
      sensitivities). } 
    \label{fig:doubleBR}
\end{figure*}

\medskip
Let us now address the question of whether a joint measurement of the
BRs and $\theta_{13}$ can shed some light on experimentally unreachable
parameters, like $m_{N_3}$. 
The expected improvement in the experimental sensitivity to the LFV
ratios supports the possibility that 
a BR could be measured in the future, thus
providing the first experimental evidence for new
physics, even before its discovery at the LHC.
The prospects are especially encouraging regarding $\mu \to e\,
\gamma$, where the experimental sensitivity will improve by at least two
orders of magnitude. Moreover, and given the impressive
effort on experimental neutrino physics, a measurement
of $\theta_{13}$ will likely also occur in the 
future~\cite{theta13_sensitivities}.  

Given that, as previously emphasised, $\mu \to e\,\gamma$ is very
sensitive to $\theta_{13}$, whereas this is not the case for 
BR($\tau \to \mu\,\gamma$), 
and that both BRs display the same approximate behaviour with 
$m_{N_3}$ and $\tan \beta$, we have studied the correlation between
these two observables. This optimises the impact of a 
$\theta_{13}$ measurement, since it allows to minimise the uncertainty
introduced from not knowing $\tan \beta$ and $m_{N_3}$, and at the
same time offers a better illustration of the uncertainty associated
with the $R$-matrix angles.
In this case, the correlation of the BRs with respect to $m_{N_3}$
means that, for a fixed set of parameters, varying $m_{N_3}$ implies
that the predicted point 
(BR($\tau \to \mu\,\gamma$),~BR($\mu \to e \, \gamma$)) 
moves along a line with approximately constant slope in the 
BR($\tau \to \mu\,\gamma$)-BR($\mu \to e \, \gamma$) plane.
On the other hand, varying $\theta_{13}$ leads to a 
displacement of the point along the vertical axis.

In Fig.~\ref{fig:doubleBR}, we illustrate this correlation for SPS
1a, choosing distinct values of the heaviest neutrino mass, and
scanning over the BAU-enabling $R$-matrix angles (setting $\theta_3$ to
zero) as 
\begin{eqnarray}\label{doubleBR:input}
& 0\, \lesssim \,|\theta_1|\,\lesssim \, \pi/4 \,, \quad \quad
-\pi/4\, \lesssim \,\arg \theta_1\,\lesssim \, \pi/4 \,, \nonumber \\
& 0\, \lesssim \,|\theta_2|\,\lesssim \, \pi/4 \,, \quad \quad
\quad \,\,\,\,\,\,
0\, \lesssim \,\arg \theta_2\,\lesssim \, \pi/4 \,, \nonumber \\
& m_{N_3}\,=\,10^{12}\,,\,10^{13}\,,\,10^{14}\,\text{GeV}\,.
\end{eqnarray} 
We considered the following values,
$\theta_{13}=1^\circ$, $3^\circ$, $5^\circ$ and $10^\circ$, and only
included in the plot the BR predictions which allow for a favourable BAU. 
Other SPS points have also been considered but they are not shown here for
brevity (see~\cite{Antusch:2006vw}).
We clearly observe in Fig.~\ref{fig:doubleBR} that  
for a fixed value of $m_{N_3}$, and for a given value of $\theta_{13}$, the
dispersion arising from a $\theta_1$ and $\theta_2$ variation produces a small
area rather than a point in the 
BR($\tau \to\mu\,\gamma$)-BR($\mu \to e \, \gamma$) plane.

The dispersion along the BR($\tau \to \mu\,\gamma$) axis is of
approximately one order of magnitude for all $\theta_{13}$. 
In contrast, the dispersion along the BR($\mu \to e\,\gamma$) axis
increases with decreasing $\theta_{13}$,
ranging from 
an order of magnitude for $\theta_{13}=10^\circ$,
to over three orders of magnitude for the case of small $\theta_{13}$
($1^\circ$). 
From Fig.~\ref{fig:doubleBR} 
we can also infer that other choices of $m_{N_3}$ (for $\theta_{13}
\in [1^\circ, 10^\circ]$) would lead to BR
predictions which would roughly lie within the diagonal lines depicted
in the plot. Comparing
these predictions for the shaded areas along the expected diagonal
``corridor'', with the allowed experimental region, allows to conclude
about the impact of a $\theta_{13}$ measurement on the allowed/excluded 
$m_{N_3}$ values.

The most important conclusion from Fig.~\ref{fig:doubleBR} is that for
SPS~1a, and for the parameter space defined in Eq.~(\ref{doubleBR:input}), 
an hypothetical $\theta_{13}$ measurement larger than $1^\circ$, together 
with the present experimental bound on the BR($\mu \to e\,\gamma$),
will have the impact of excluding values of $m_{N_3} \gtrsim 10^{14}$
GeV. Moreover, with the planned MEG
sensitivity, the same $\theta_{13}$ measurement can further constrain 
$m_{N_3} \lesssim 3\times 10^{12}$~GeV.
The impact of any other $\theta_{13}$ measurement can be analogously
extracted from Fig.~\ref{fig:doubleBR}.

\bigskip
As a final comment let us add that, remarkably, 
within a particular SUSY scenario and scanning over 
specific $\theta_1$ and $\theta_2$ BAU-enabling ranges for various 
values of $\theta_{13}$, the comparison of the
theoretical predictions for BR($\mu \to e\,\gamma$) and 
BR($\tau \to \mu\,\gamma$) with the present experimental bounds allows 
to set $\theta_{13}$-dependent upper bounds on $m_{N_3}$. 
Together with the indirect lower bound arising from leptogenesis 
considerations, this clearly provides interesting hints on the value of the 
seesaw parameter $m_{N_3}$.
With the planned future sensitivities, these bounds would further improve by
approximately one order of magnitude.

Ultimately, a joint measurement of the LFV branching ratios, 
$\theta_{13}$ and the sparticle spectrum would be a powerful tool for 
shedding some light on otherwise unreachable SUSY seesaw parameters.
It is clear from all this study that the interplay between LFV processes and
future improvement in neutrino data is challenging for the searches of
new physics.   

\bigskip
\begin{acknowledgments}
A.~M.~Teixeira is grateful to A.~Abada for her help in preparing this
presentation. This work has been  
supported by the French ANR project PHYS@COL\&COS. 
\end{acknowledgments}

\end{document}